\documentclass[12pt]{article}
\usepackage{epsfig}

\def\beq{\begin{equation}}
\def\eeq#1{\label{#1}\end{equation}}
\def\eeqn{\end{equation}}

\def\beqa{\begin{eqnarray}}
\def\eeqa#1{\label{#1}\end{eqnarray}}
\def\eeqan{\end{eqnarray}}

\let\bar=\overbar

\def\Dslash{\not{\hbox{\kern-4pt $D$}}}
\def\dslash{\not{\hbox{\kern-2pt $\del$}}}

\def\msb{{\bar{\ssstyle M \kern -1pt S}}}

\def\Title#1{\begin{center} {\Large {\bf #1} } \end{center}}

\begin{document}

\Title{Confinement and dual superconductivity of QCD vacuum.}

\bigskip\bigskip

%+\addtocontents{toc}{{\it D. Reggiano}}
%+\label{ReggianoStart}

\begin{raggedright}  

{\it Adriano Di Giacomo\index{Di Giacomo, A.}\\
Department of Physics Pisa University and INFN\\
3  Largo B. Pontecorvo\\
I-56127 Pisa, ITALY}
\bigskip\bigskip
\end{raggedright}

\section{Introduction}

Quarks are confined in Nature. The ratio of the quark abundance $n_q$ to that of the nucleons $n_p$
has the experimental upper bound     ${n_q \over n_p} \le 10^{-27} $  ~\cite{PDG} to be compared to the expectation in the absence of confinement in the Standard Cosmological Model ${n_q \over n_p} \approx 10^{-12} $  ~\cite{Okun}.  The inclusive cross section for production of quarks + antiquarks  in $p + p$ collisions  $\sigma_q \equiv \sigma (p+p \to q(\bar q) + X )$ has the upper bound  $\sigma_q \le 10^{-40} cm^2$ ~\cite{PDG} to be compared to the expectation in perturbative $QCD$  $\sigma_q \approx \sigma_{total} \approx 10^{-25} cm^2$.   The inhibition factor is, in the two  cases $\le 10^{-15}$ which is a very small number.  The only natural explanation is that  both $n_q$ and $\sigma _q$ are strictly zero due to some symmetry.  

If this is true the deconfining transition is a change of symmetry, i.e. an order disorder transition, and cannot be a continuous crossover. A crossover would request a fine tuning of 15 orders of magnitude which is non-natural .  This is well established  in pure gauge theory (no quarks) , where the Polyakov line is the order parameter. It is controversial in the presence of light quarks. I will briefly review the status of this statement.

A candidate symmetry for confinement is Dual Superconductivity of vacuum~\cite{'tHooft2},~\cite{Man}. The idea is that in the confined phase , ($T \le T_c$), there is Higgs breaking of some magnetic symmetry, and a dual Meissner effect which channels the chromoelectric field acting between a $q - \bar q$ pair into an Abrikosov flux tube , whose energy is proportional to the distance. Above the transition ($T  > T_c$)  magnetic symmetry is restored and with it deconfinement.

The confining vacuum in this model , due to monopole condensation,  is a superposition of states with different magnetic charge, or a Bogolubov-Valatin vacuum.  Above $T_c$ the vacuum becomes normal and magnetic charge is super-selected.

 Two alternative strategies have been developed to investigate this phenomenon :
 
 1) Look at the symmetry~\cite{Zak},~\cite{DelDigPaf1}~\cite{DelDigPaf2}~\cite{Frolich} .
 The vacuum expectation value ($vev$) of a gauge invariant  operator $\mu$ carrying non zero magnetic charge , $\langle \mu \rangle$ ,  can  be an order parameter

    $ \langle \mu \rangle \neq 0 $       for        $ T <  T_c $       (confined)

   $ \langle \mu \rangle = 0 $             for         $ T \ge T_c $    (deconfined)
 
 2) Expose monopoles in lattice configurations in some gauge ( usually the so called maximal-abelian gauge~\cite{Suzuki}),  look for monopole dominance and try to extract a monopole effective action, out of which  condensation can eventually be read.
 
 I will briefly review the status of these two approaches.
 
\section{$N_f=2$ QCD : first order or crossover?}

The chiral transition at  $m_q=0$ can be analyzed by use of  $4 -\epsilon$ plus renormalization group techniques ~\cite{PisWil} .  For $N_f \ge 3$ there is no infrared stable fixed point, and therefore the transition is first order. For $N_f=2$ instead there are two possibilities:

a) The transition is second order in the universality class $O(4)$ , and then at small values of the mass it is a crossover. A tricritical point is expected at the some value of the mass, where the transition becomes again first order.

b) The transition is first order and then it is first order also at non zero masses. No tricritical point needed.

The scenario  a)  excludes a change of symmetry at the transition, and points to an "unnatural" choice 
of Nature, in the language of  Sect.1.  

The scenario  b)  is compatible with a change of symmetry at the deconfining transition.

In principle the existence of the tricritical point can be established by heavy ion experiments. Up to now there is no indication of its existence.

The problem can be settled by use of lattice simulations measuring the critical indexes of the transition, e.g. by finite size scaling analysis.
However the problem proves to be difficult  mainly due to the presence of two independent scaling variables. The pioneering papers on the subject are rather inconclusive~\cite{Bie}~\cite{Fuk}~\cite{Mil}.
 Due to the fundamental importance of the problem we are carrying out a systematic research program
 on it with an unprecedented effort~\cite{DelDigPic1}~\cite{DelDigPic2}. I will summarize the present status of the program.
 
 Denoting by $L_s$ the spatial size of the lattice, the general scaling laws hold for the specific heat $C_V$ and for the susceptibility of the order parameter $\chi$
 \begin{eqnarray}
 C_V - C_0 = L_s^{\alpha \over \nu} \Phi_c (\tau L_s^{1\over \nu}, m L_s^{y_h})\\
 \chi - \chi_0 = L_s^{1\over \nu} \Phi_{\chi} (\tau L_s^{1\over \nu}, m L_s^{y_h})  
 \end{eqnarray}
 Here  $C_0, \chi_0$ are ultraviolet subtractions ,$\tau \equiv 1 - {T \over T_c}$ is the reduced temperature, and the critical indexes $\alpha $, $\gamma $, $\nu $, $y_h $ characterize the universality class of the transition . 
 
 For second order $O(4)$  $\alpha= -.24, \gamma=1.48, \nu=.74, y_h =2.49$
 
 For weak first order  $\alpha= 1, \gamma=1, \nu=.{1\over 3}, y_h =3$.
 
 Our strategy has been to keep one of the scaling variables  fixed in the scaling laws Eq's(1)(2) , and to check the scaling with respect to the other.  To keep e.g. the second variable fixed one has to assume a value for 
 $y_h$ , either the value $y_h=2.49$ corresponding to $O(4)$ [or $O(2)$]  , or the value $y_h =3$ corresponding to first order ,  and change  $m$ and $L_s$  in such a way that $m L_s^{y_h}$ stays unchanged. One then verifies if the scaling in the other variable is  consistent with the choice of universality class , [Eq(1) with $m L_s^{y_h}$ fixed].
 This has been done in ~\cite{DelDigPic1} assuming second order $O(4)$ and this choice proved to be inconsistent, and in ~\cite{DelDigPic2} assuming first order with a positive result.  The result is illustrated in Fig. (1) 
 \begin{figure}[htb]
\begin{center}
  \includegraphics[width=0.49\textwidth,clip=]{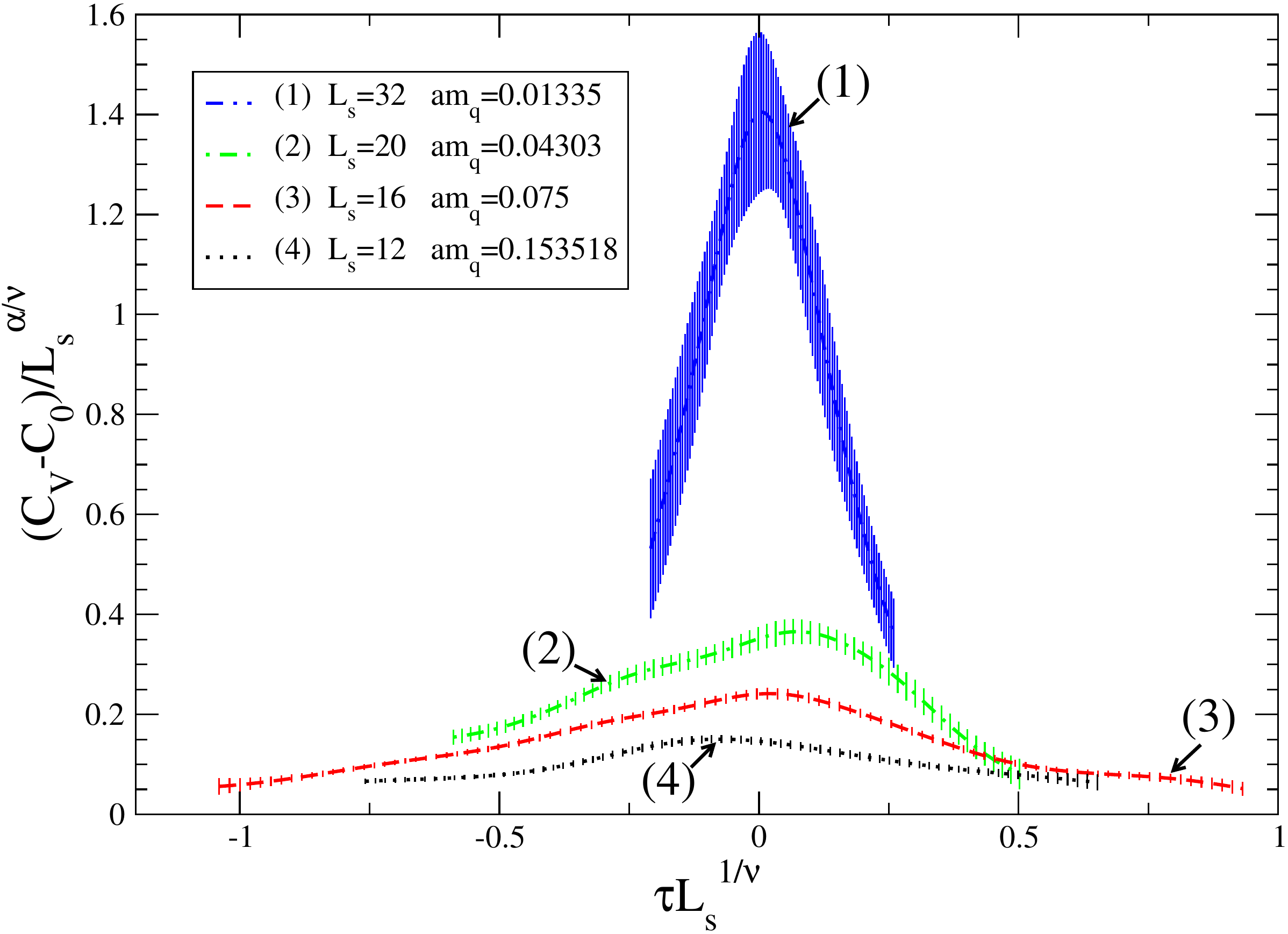}\includegraphics[width=0.51\textwidth,clip=]{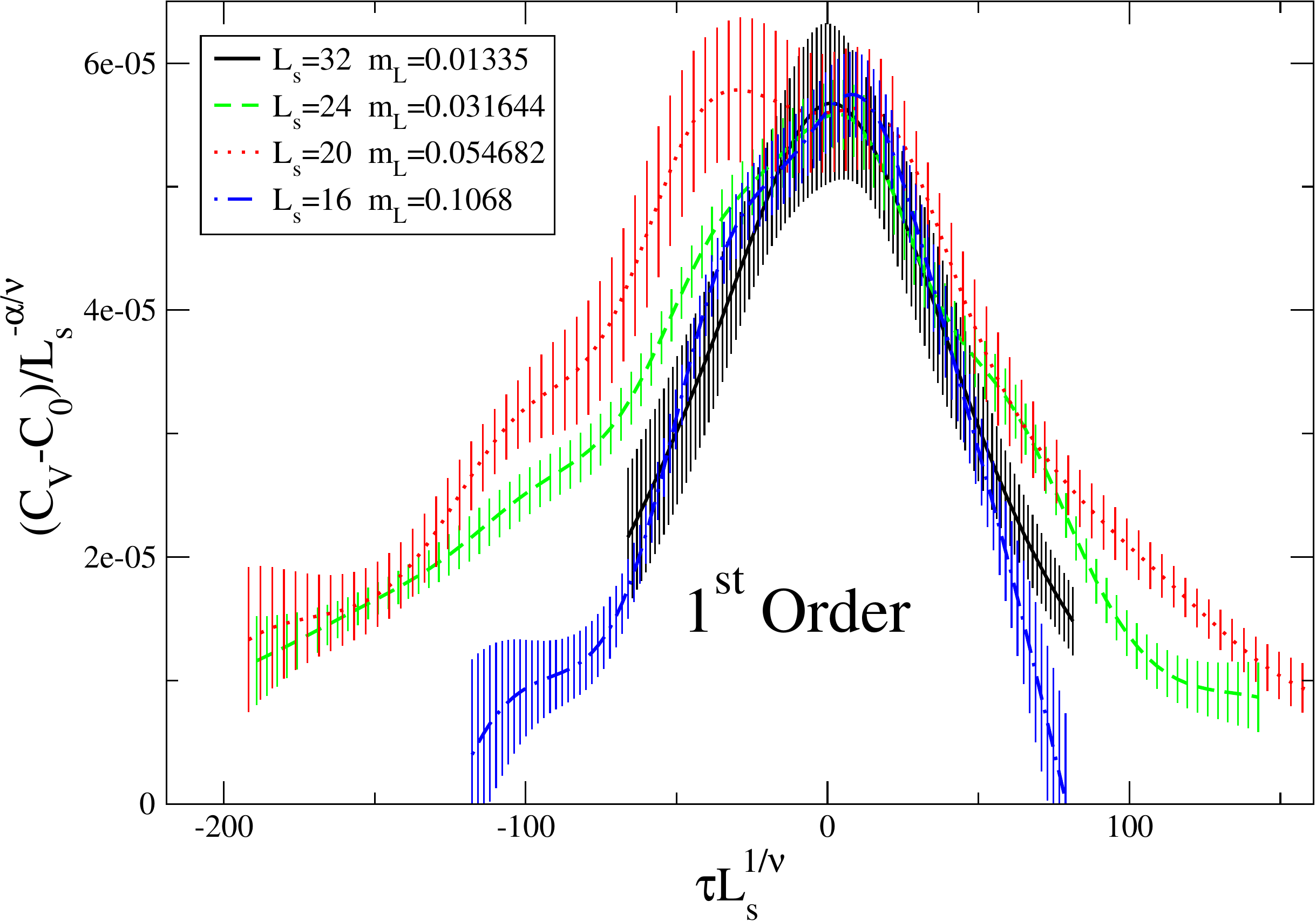}
\caption{Scaling Eq.(1) assuming O(4) [left]  and first order [right]. If the scaling is correct the curves for different $L_s$ should coincide .}
\label{fig1}
\end{center}
\end{figure}

Alternatively one can keep the  variable  $\tau L_s^{1\over \nu}$ fixed and look at the dependence on
the other one. If the chiral transition is second order everything should be analytic as $L_s \to \infty$
and this implies the scaling laws
\begin{eqnarray}
C_V - C_0 \approx_{L_s \to \infty}  m^{\alpha \over {\nu y_h}}  f_C(\tau L_s^{1\over \nu})\\
\chi - \chi_0\approx_{L_s \to \infty} m^{\gamma \over {\nu y_h}}  f_{\chi}(\tau L_s^{1\over \nu})
\end{eqnarray}
For first order these equations have an additional term proportional to the volume $L_s^3 f_1(\tau L_s^{1\over \nu})$, which makes both the susceptibilities divergent as the volume goes  large.
For a weak first order the additional term becomes dominant only at large volumes, and is strongly peaked at $T= T_c$ as expected from  a discontinuity of the internal energy.  At present spatial volumes 
(up to $32^3$ spatial lattices), the diverging term is still small compared to the first one~\cite{DelDigPic2}.
\begin{figure}[htb]
  \begin{center}
    \includegraphics[width=0.49\textwidth,clip=]{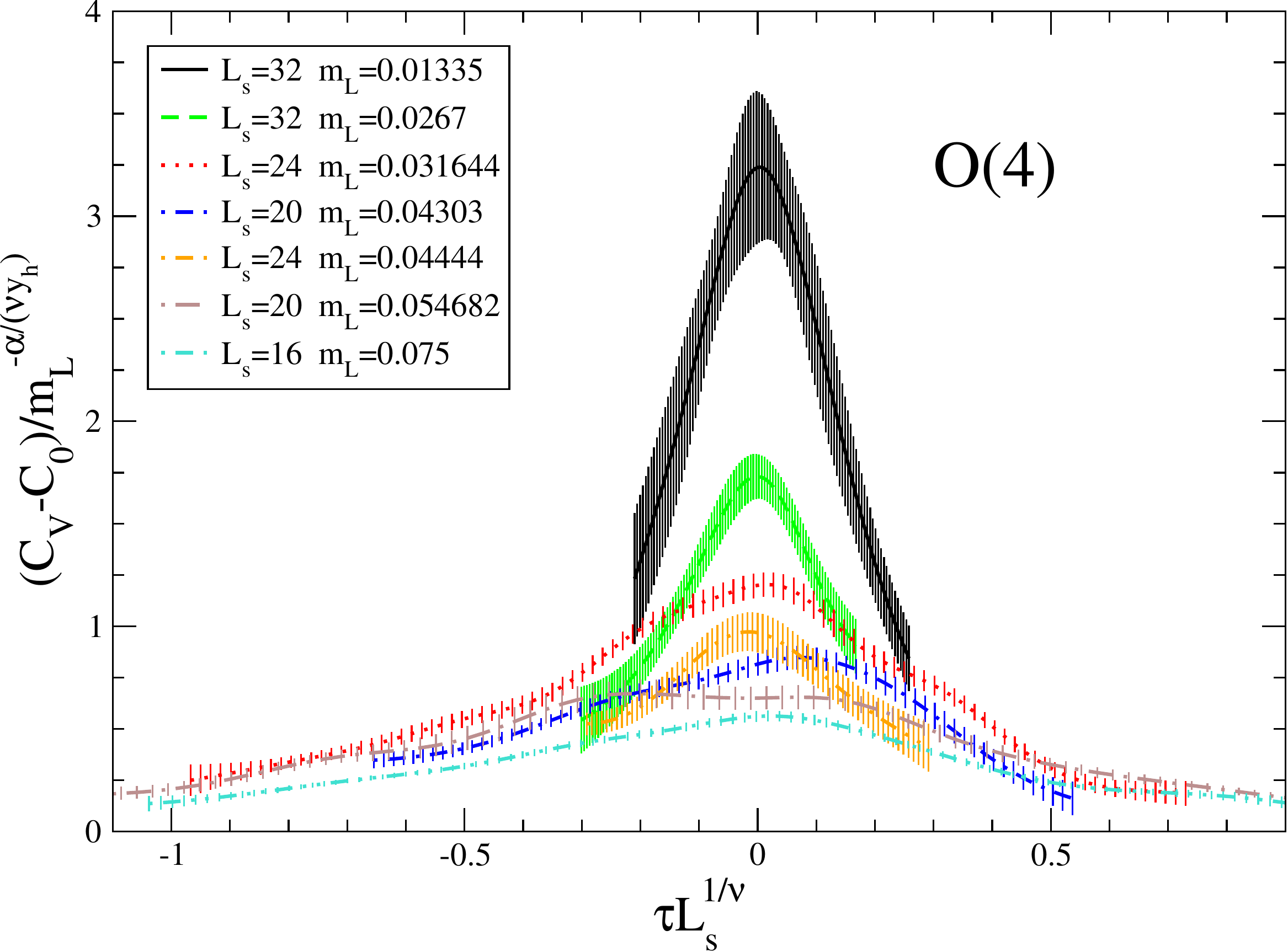}\includegraphics[width=0.51\textwidth,clip=]{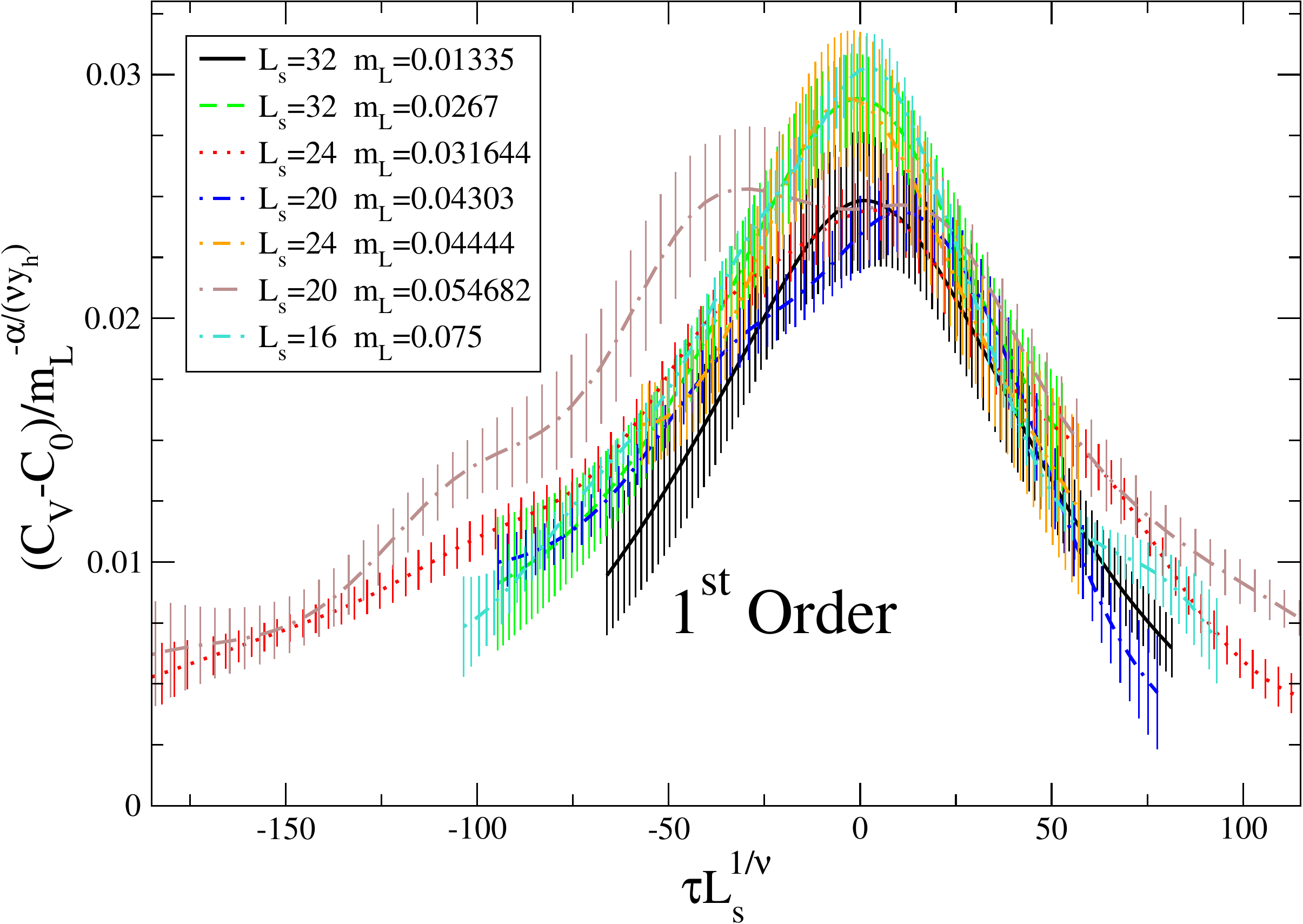}
    \caption{Scaling Eq.(3) assuming O(4) [left]  and first order [right]. If the scaling is correct the curves for different $L_s$ should coincide .}
    \label{fig2}
  \end{center}
\end{figure}

In Fig(2) the scaling Eq(3) is shown for second order $O(4)$ and for first order. The first possibility is definitely excluded. First order is consistent, but it will possibly be confirmed by simulations at larger volumes, where the diverging part will show up.  Notice that we mainly rely on the specific heat, which is independent on any prejudice on the choice of the order parameter.
\section{Dual superconductivity as a mechanism for confinement.}
The idea goes back to ~\cite{'tHooft2},~\cite{Man} . If the vacuum is a condensate of magnetic charges
a dual Meissner effect takes place which channels the chromoelectric field acting between a $q-\bar q$
pair into an Abrikosov flux tube, whose energy is proportional to the distance.  The effect disappears above the deconfining temperature $T_c$ , where the magnetic charge of the vacuum has a definite value.
Two alternative strategies have been used to detect this effect:

1)  Look at the symmetry.  The Higgs breaking of the magnetic gauge symmetry is detected by measuring the $vev$ of an operator $\mu$ carrying magnetic charge .  Below $T_c$ $\langle \mu \rangle \neq 0$ , above $T_c$ $\langle \mu \rangle = 0$ in the thermodynamical limit $V \to \infty$.
The operator has been developed and tested on a number of known systems~\cite{DigLucMon1},~\cite{DigLucMon2}~\cite{Car1}~\cite{Car2}~\cite{DelDigLuc}.

2) Expose monopoles in lattice configurations in some gauge ( Maximal Abelian Gauge)~ \cite{Suzuki}
look  for monopole dominance and try to extract from lattice data a monopole effective action, in which hopefully  condensation can be read.

About the approach  2)  I will only quote a recent work ~\cite{Suz2} in which monopole dominance, which was usually considered a specific property of the Maximal Abelian Gauge, is instead shown to be a general property
of monopoles in any gauge. The parallel property in the approach  1)  was known since long time ~\cite{Car1}~\cite{CeaCos}.

The approach 1) is by now well established. The operator $\mu$ is defined as~\cite{DigPaf}~\cite{DigLucMon1}~\cite{DigLucMon2}~\cite{Frolich}
\begin{equation}
\mu (\vec x , t) = \exp [{ i q {1\over g^2}\int d^3y \vec E(\vec y, t)\vec b_{\perp}(\vec y - \vec x) }]
\end{equation}
In Eq(5) only the transverse component of the electric field survives the convolution, i.e. the conjugate momentum to $\vec A_{\perp}$ , so that the operator $\mu$ shifts  the value of  $\vec A_{\perp}$ by  $\vec b_{\perp}$, which is chosen to be the vector potential produced by a monopole sitting in $\vec x$
in the transverse gauge.  $\mu$ creates a monopole with $q$ units of the Dirac-quantised magnetic charge.
If vacuum has a definite magnetic charge $\langle \mu \rangle =0$;  $\langle \mu \rangle \neq 0$ signals dual superconductivity.

Instead of  $\langle \mu \rangle $ it proves convenient to use the susceptibility  $\rho \equiv {{\partial \ln(\langle \mu \rangle )}\over{\partial \beta}}$ where $\beta = {{2N}\over g^2}$ ~\cite{DelDigPaf1}. $\rho$ obeys the scaling law 
\begin{equation}
\rho /L_s^{1\over \nu} = \Phi(\tau L_s^{1\over \nu} )
\end{equation}
We have analyzed $U(1)$~\cite{DigPaf}, $SU(2)$~\cite{DigLucMon1}, $SU(3)$~\cite{DigLucMon2}  pure gauge theories and  $N_f=2$ $QCD$~\cite{DDLP}. In all cases  $\langle \mu \rangle $ proves to be a good order parameter . This result is a theorem for $U(1)$ gauge theory\cite{Frolich}. For $SU(2)$ and $SU(3)$ the scaling law Eq.(6) is obeyed with the critical index $\nu$ of
3d-Ising and first order respectively, in agreement with the determination from the Polyakov line.
For $N_f=2$ $QCD$ the Polyakov line is not an order parameter, but  $\langle \mu \rangle $ is, and  the scaling Eq(6) is consistent with a first order transition~\cite{DDLP}.

\section{Conclusions}
The deconfining transition is  an order disorder transition in the systems we have studied.
There is evidence that dual superconductivity is the symmetry behind confinement.

%%%%%%%%%%%%%%%%%%%%%%%%%%%%%%%%%%%%%%%%%%%%%%%%%%%%%%%%%%%%%%%%%%%%%%%%%
%%
%%   use this format to include an .eps figure into your paper
%%
%%%%%%%%%%%%%%%%%%%%%%%%%%%%%%%%%%%%%%%%%%%%%%%%%%%%%%%%%%%%%%%%%%%%%%%%%%%

%%%%%%%%%%%%%%%%%%%%%%%%%%%%%%%%%%%%%%%%%%%%%%%%%%%%%%%%%%%%%%%%%%%%%%%%%
%%
%%   use this format to include a LaTeX table  into your paper
%%
%\begin{table}[b]
%\begin{center}
%\begin{tabular}{l|ccc}  
%Patient &  Initial level($\mu$g/cc) &  w. Magnet &  
%w. Magnet and Sound \\ \hline
 %Guglielmo B.  &   0.12     &     0.10      &     0.001  \\
 %Ferrando di N. &  0.15     &     0.11      &  $< 0.0005$ \\ \hline
%\end{tabular}
%$\caption{Blood cyanide levels for the two patients.}
%\label{tab:blood}
%\end{center}
%\end{table}
%%%%%%%%%%%%%%%%%%%%%%%%%%%%%%%%%%%%%%%%%%%%%%%%%%%%%%%%%%%%%%%%%%%%%%%%%%%

\bigskip
I am grateful to G. Cossu, M. D'Elia, B. Lucini, G. Paffuti, C. Pica who are the collaborators in 
this investigation.

\end{document}